\documentclass[aps,prl,twocolumn,superscriptaddress,amsmath,amssymb,floatfix,reprint]{revtex4-2}

\usepackage{CJK}
\usepackage{newtxtext}
\usepackage{graphicx}
\usepackage{bm}
\usepackage{siunitx}
\usepackage{derivative}
\usepackage[nopartial]{fixdif}
\usepackage{physics2}
\usephysicsmodule{ab, ab.legacy}
\definecolor{linkblue}{RGB}{40,53,142}
\usepackage{hyperref}
\hypersetup{
    colorlinks=true,
    linkcolor=linkblue,
    urlcolor=linkblue,
    citecolor=linkblue,
    pdfpagemode=UseNone,
    linkbordercolor=0 1 0,
    pdfstartview=
}
\usepackage{soul}
\usepackage[]{cleveref}
\crefname{figure}{Fig.}{Figs.}
\Crefname{figure}{Figure}{Figures}
\crefname{table}{table}{tables}
\Crefname{table}{Table}{Tables}
\crefname{equation}{Eq.}{Eqs.}
\Crefname{equation}{Equation}{Equations}
\crefname{section}{Sec.}{Secs.}
\Crefname{section}{Section}{Sections}
\crefname{subsection}{Sec.}{Sec.}
\crefrangeformat{equation}{(#3#1#4)--(#5#2#6)}

\newcommand{\grad}{\bm{\nabla}}
\newcommand{\divergence}{\bm{\nabla}\cdot}
\newcommand{\laplacian}{\nabla^2}

\newcommand{\kBT}{k_\mathrm{B}T}
\newcommand{\bx}{\bm{x}}

\newcommand{\bk}{\bm{k}}
\newcommand{\bu}{\bm{u}}
\newcommand{\buhat}{\hat{\bu}}
\newcommand{\bv}{\bm{v}}

\newcommand{\bP}{\bm{\mathcal{P}}}
\newcommand{\W}{\mathcal{W}}
\newcommand{\bW}{\bm{\mathcal{W}}}

\newcommand{\cZ}{\mathcal{Z}}
\newcommand{\bcZ}{\bm{\cZ}}

\newcommand{\corr}{C}
\newcommand{\corrhat}{\widehat{C}}
\newcommand{\autocorr}{C_{v}}
\newcommand{\bX}{\bm{X}}
\newcommand{\nueff}{\nu_{\text{eff}}}
\newcommand{\nueffhat}{\hat{\nu}_{\text{eff}}}

\begin{document}

\title{
Spatially Correlated Noise Induces Transitions from the Diffusive to Ballistic Regime in Fluids
}

\begin{CJK*}{UTF8}{gbsn}
\author{Sijie Huang (黄斯杰)}
\author{Ayush Saurabh}
\affiliation{
Department of Physics, Arizona State University, Tempe, AZ 85287, USA
}
\affiliation{
Center for Biological Physics, Arizona State University, Tempe, AZ 85287, USA
}
\author{Steve Press\'e}
\email{Corresponding author: spresse@asu.edu}
\affiliation{
Department of Physics, Arizona State University, Tempe, AZ 85287, USA
}
\affiliation{
Center for Biological Physics, Arizona State University, Tempe, AZ 85287, USA
}
\affiliation{
School of Molecular Sciences, Arizona State University, Tempe, AZ 85287, USA
}

\date{\today}

\begin{abstract}
We investigate the fluctuating incompressible Navier--Stokes equation driven by spatially correlated thermal noise characterized by a single length scale. This formulation is constructed to preserve thermal equilibrium through the fluctuation--dissipation relation (FDR), which enforces the same spatial correlation in the viscous diffusion term and therefore gives rise to nonlocal momentum transport. Numerical simulations of tracer diffusion in fluids governed by this formulation reveal that the mean-squared displacement (MSD) depends monotonically on the correlation length \(\ell\) and the correlation strength \(\beta\). Intuitively, increasing \(\ell\) enhances MSD and induces the emergence of an early-time ballistic regime, as a larger correlation length slows momentum diffusion. Counterintuitively, decreasing \(\beta\) also increases the MSD, since a weaker correlation strength also retards momentum diffusion, whereas smaller \(\ell\) or larger \(\beta\) suppresses the ballistic regime and leads to a diffusive behavior. The emergence or suppression of the ballistic regime stems from how spatial correlations, incorporated through the FDR to maintain equilibrium, alter the effective momentum transport across scales. Interestingly, we further show that the resulting nonlocal diffusion is reminiscent of the slow dynamics in glassy and other disordered systems.
\end{abstract}

\maketitle

\end{CJK*}

Deviations of the mean-squared displacement~(MSD) from the linear time scaling of Brownian motion, termed anomalous diffusion, are encountered across a range of fields, from biophysics and soft matter to astrophysics and finance~\cite{christov2012pnas,jiang2019rpp,viswanathan1996nature,sabri2020prl,graham2022nc}. Anomalous diffusion often arises under nonequilibrium conditions, such as active matter, living cells, or externally driven systems~\cite{dieterich2008pnas,mukherjee2021prl}. In these systems, continuous external driving gives rise to spontaneously emerging spatial correlations---a hallmark of nonequilibrium dynamics---such as in hydrodynamic turbulence~\cite{richardson1926,mukherjee2021prl,singh2024nc}. However, even in equilibrium systems, anomalous diffusion can arise from intrinsic structural correlations rather than external driving; for example, through geometric confinement, structural disorder, or viscoelastic memory~\cite{rusciano2022prl,vinales2020pre,wang2009pnas}. This motivates our exploration of whether intrinsic noise correlations alone can modify diffusion dynamics at equilibrium. 

While the examples above highlight the diverse physical origins of anomalous diffusion, fluctuating hydrodynamics provides a simple theoretical framework to examine this behavior systematically. In its standard form, intrinsic thermal fluctuations enter the incompressible Navier--Stokes equation as a random stress tensor modeled as spatiotemporal white noise. This formulation yields the full hydrodynamic description of immersed particles undergoing classical Brownian motion~\cite{zwanzig1964,hauge1973jsp,bedeaux1974,usabiaga2013jchemphys}. At the hydrodynamic level, anomalous diffusion is often introduced phenomenologically through externally imposed correlations that mimic heterogeneity or confinement~\cite{liu2024jfm,palmer2020pre,spakowitz2019,pierro2018pnas,tyukhova2016}. In contrast, the role of intrinsic spatial correlations within fluctuating hydrodynamics remains largely unexplored. Such intrinsic correlations could arise, for example, from finite-range stress propagation in structured or viscoelastic fluids~\cite{ma2021pre,behbahani2024prr,lau2003prl,klochko2018sm,grimm2025sm}, where nonlocal stress responses may give rise to spatially correlated thermal fluctuations.

In this Letter, we investigate how intrinsic spatial correlations in thermal noise affect tracer diffusion in fluids at thermal equilibrium. We introduce a fluctuating incompressible Navier--Stokes equation driven by spatially correlated noise characterized by a single correlation length \(\ell\). To preserve equilibrium, we derive and enforce the corresponding fluctuation--dissipation relation (FDR). The FDR requires that when thermal noise is spatially correlated, viscous momentum diffusion becomes nonlocal, extending over finite distances and thereby acquiring a scale dependence. This nonlocal diffusion resembles the convolution kernels used in biological and turbulent systems to describe aggregation, migration, and scale-dependent transport~\cite{painter2024,chen2019,morale2005jmb,ha2008krm,liu2023prf,hamba2022jfm}.

Numerical simulations of tracer particles diffusing in the fluids governed by the spatially correlated fluctuating Navier--Stokes equation reveal that the MSD depends monotonically on both the correlation length \(\ell\) and strength \(\beta\). Increasing \(\ell\) or decreasing \(\beta\) enhances the MSD and induces an emergence of the ballistic regime, while smaller \(\ell\) or larger \(\beta\) suppresses it. The emergence or suppression of ballistic behavior reflects how spatial correlations alter momentum diffusion across scales, either enhancing or hindering the effective momentum transport. In particular, the slowdown of momentum diffusion mirrors the sluggish dynamics of glassy systems, suggesting a conceptual link between equilibrium correlated hydrodynamics and glass-like behavior~\cite{pastore2014sm,chaudhuri2007prl,weeks2000science}. 

\begin{figure*}
    \centering
    \includegraphics[width=\linewidth]{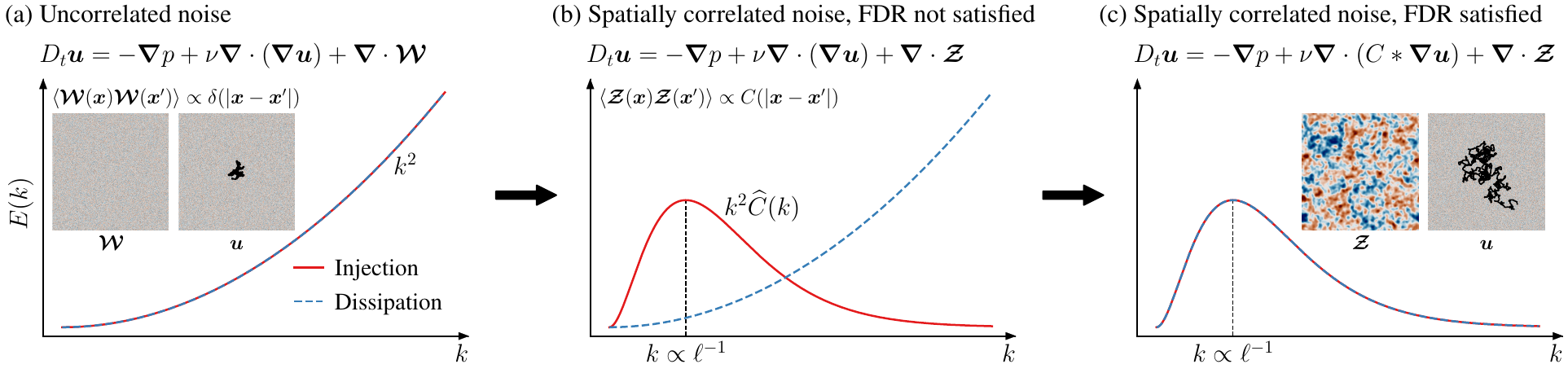}
    \caption{FDR requires a scale-by-scale balance between noise-energy injection and viscous dissipation. (a) For uncorrelated noise, both noise-energy injection and viscous dissipation scale as \(k^2\), corresponding to the Laplacian, and thus balance each other. The material derivative is \(D_t=\partial_t + \bu\cdot\grad\). Insets show instantaneous noise and velocity fields, both spatially random, consistent with thermal equilibrium. The black line in the velocity field represents a sample particle trajectory. (b) Introducing spatial correlations with length scale \(\ell\) redistributes the noise-energy injection across scales. If the viscous term remains unmodified, the balance between energy injection and dissipation is lost. (c) Incorporating the same spatial correlation into the viscous term restores the scale-by-scale energy balance. In the insets, the correlated noise field develops large-scale structures, whereas the velocity field remains random, confirming thermal equilibrium. The particle trajectory exhibits enhanced diffusion compared with (a).}
    \label{fig:1}
\end{figure*}

\textit{Formulation Outline}---To derive the spatially correlated fluctuating Navier--Stokes equation, we start with the standard fluctuating hydrodynamics~\cite{landau1959} 
\begin{equation}
    \label{eq:llns}
    \partial_t\bu + \bu\cdot\grad\bu = -\grad p + \nu\laplacian\bu + \sqrt{2\nu\Theta}\divergence\bW,
\end{equation}
which describes thermally fluctuating fluids at the continuum level. Here, \(\bu\) is the fluid velocity field, \(p\) the pressure, $\nu$ the kinematic viscosity, and the noise amplitude \(\Theta = \kBT/\rho\) is proportional to the thermal energy at temperature $T$ and density $\rho$. The velocity field \(\bu\) is interpreted as a coarse-grained quantity, defined on length scales larger than the molecular mean free path~\cite{bandak2022,bandak2024}. For incompressible, isothermal fluids, \(\rho\), \(\nu\), and \(T\) are taken as constants. The random stress \(\bW\) models the thermal noise as a white Gaussian field with covariance~\cite{zarate2006}
\begin{align}
    \label{eq:uncorrelated_noise}
    \aab{\W_{ij}(\bx,t)\W_{lm}(\bx',t')} &= (\delta_{il}\delta_{jm} + \delta_{im}\delta_{jl}) \nonumber \\ &\hspace{10pt}\times\delta(r)\delta(t-t'),
\end{align}
where \(r=|\bx-\bx'|\) is the spatial separation. 

The fluids governed by \cref{eq:llns} are in thermal equilibrium because the noise-energy injection and viscous dissipation balance at every scale~[\cref{fig:1}(a)]. In Fourier space, the noise-energy injection scales as \(k^2\), consistent with the Laplacian operator, ensuring that energy injected by thermal fluctuations is exactly dissipated by viscosity in accordance with the FDR.

We generalize \cref{eq:llns} by replacing the spatiotemporal white noise \(\bW\) with a spatially correlated, temporally white Gaussian field \(\bcZ\), whose covariance is 
\begin{align}
    \label{eq:correlated_noise}
    \aab{\cZ_{ij}(\bx,t)\cZ_{lm}(\bx',t')} &= (\delta_{il}\delta_{jm} + \delta_{im}\delta_{jl})\nonumber\\ &\hspace{10pt}\times\corr(r)\delta(t-t').
\end{align}
The correlation function \(\corr(r)\) depends only on the separation \(r\) and is characterized by a single length scale \(\ell\). 

Spatial correlations redistribute noise-energy injection across scales, breaking thermal equilibrium by unbalancing energy injection and viscous dissipation~[\cref{fig:1}(b)]. To restore equilibrium, the FDR requires that the viscous diffusion term incorporate the same spatial correlation, thereby re-establishing the scale-by-scale balance~[\cref{fig:1}(c)]. The FDR can be derived in Fourier space (see End Matter for a summary and the Supplemental Material~\cite{supplemental} for the full derivation), which yields the spatially correlated fluctuating Navier--Stokes equation with a nonlocal diffusion term,
\begin{equation}
    \label{eq:generalized_fns}
    \partial_t\bu + \bu\cdot\grad\bu = -\grad p + \divergence(\nueff*\grad\bu) + \sqrt{2\nu\Theta}\divergence\bcZ,
\end{equation}
where \(\nueff(r) = \nu C(r)\) is an effective viscosity, and \(*\) denotes convolution. The convolution in the viscous term indicates that momentum diffusion becomes nonlocal: the viscous response at a point depends on \(\grad\bu\) in a surrounding region, reflecting spatially extended hydrodynamic coupling.

To explore how spatial correlations affect particle diffusion, we consider a two-dimensional Lorentzian-type correlation function, chosen for analytical tractability and computational efficiency. The correlation function is
\begin{equation}
    \corr_1(r) = \frac{\ell}{2\pi(\ell^2 + r^2)^{3/2}}.
\end{equation}
It approaches the delta function as \(\ell\to0\), allowing systematic deviations from the white-noise limit. As \(\ell\) increases, the correlation strength \(\corr_1(0)\propto\ell^{-2}\) decreases, reducing the overall noise amplitude while extending its spatial range. 

However, variations in the correlation length and strength of \(\corr_1\) are coupled, both influence particle diffusion simultaneously. To disentangle their individual contributions, we introduce a second correlation function
\begin{equation}
    \corr_2(r) = \frac{\beta\ell^3}{(\ell^2 + r^2)^{3/2}},
\end{equation}
where \(\beta\) fixes the overall correlation strength. In the resulting noise field, increasing \(\ell\) produces larger-scale structures, whereas increasing \(\beta\) enhances the amplitudes. Because the FDR couples forcing and dissipation, any modification of \(\corr_{1,2}\) alters both the noise-energy injection and viscous dissipation. Additional technical discussion and spectral analysis are provided in the End Matter.

\begin{figure}
    \centering
    \includegraphics[width=0.48\textwidth]{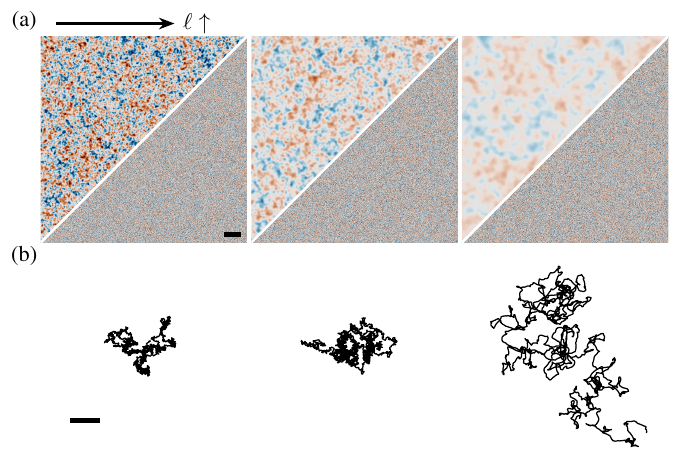}
    \caption{Enhancement of particle diffusion as the noise correlation length \(\ell\) increases for the correlation function \(\corr_1\). Particle radius \(a=\qty{0.05}{\um}\). (a) Representative instantaneous snapshots of the correlated noise (upper halves) and the resulting fluid velocity (lower halves) for \(\ell=0.05,0.1\) and \(\qty{0.2}{\um}\) (from left to right). Only the components in the \(x\)-direction are shown (\(\cZ_{xx}\) and \(u_x\)) thanks to spatial isotropy. Scale bar: \qty{1}{\um}. (b) Representative tracer trajectories over the same time span. Scale bar: \qty{0.02}{\um}. Simulation details are described in Sec.~II of the Supplemental Material~\citep{supplemental}.}
    \label{fig:2}
\end{figure}

\textit{Simulations}---We numerically solve the coupled fluid--particle system in a doubly periodic domain. Simulation details are provided in Sec.~II of the Supplemental Material~\cite{supplemental}. Briefly, \cref{eq:generalized_fns} is discretized using a Fourier--Galerkin pseudospectral scheme~\cite{canuto2007}. Particles are treated as passive, inertialess tracers advected by the fluctuating velocity field. Periodic boundaries ensure that the FDR holds without additional boundary corrections~\cite{usabiaga2012mms,delong2014jcp,atzberger2006}. Simulations systematically vary \(\ell\) and \(\beta\) for \(\corr_1\) and \(\corr_2\) to examine their effects on particle diffusion. Validation confirms energy conservation in the inviscid limit and thermal equilibrium, as evidenced by energy equipartition, Boltzmann-distributed velocities, and recovery of classical Brownian diffusion (see Figs.~S2--S4 in the Supplemental Material~\cite{supplemental}). These results demonstrate that the simulations capture both the correct equilibrium fluid state and the hydrodynamic coupling between the fluid and tracer particles.

\nocite{tamm2018prl,oksendal2003,canuto2006,donev2014jsm,cufinufft,delong2013pre}

\textit{From Diffusive to Ballistic}---\Cref{fig:2}(a) shows instantaneous fields of the correlated noise and resulting velocity for three correlation lengths, \(\ell=0.05,0.1\), and \qty{0.2}{\um}. As \(\ell\) increases, the noise develops larger-scale structures while its amplitude, set by \(\corr_1(0)\propto\ell^{-2}\), decreases. In contrast, the velocity field remains spatially random, consistent with the mode-by-mode energy equipartition. Although the snapshots appear random, the correlated noise alters the dynamics by changing how quickly fluid momentum diffuses across scales through the effective viscosity \(\nueff(r)\). This effect is not visible in instantaneous fields but manifests in the tracer statistics below and is analyzed in detail in the \textit{Discussion}.

\begin{figure}
    \centering
    \includegraphics[width=0.48\textwidth]{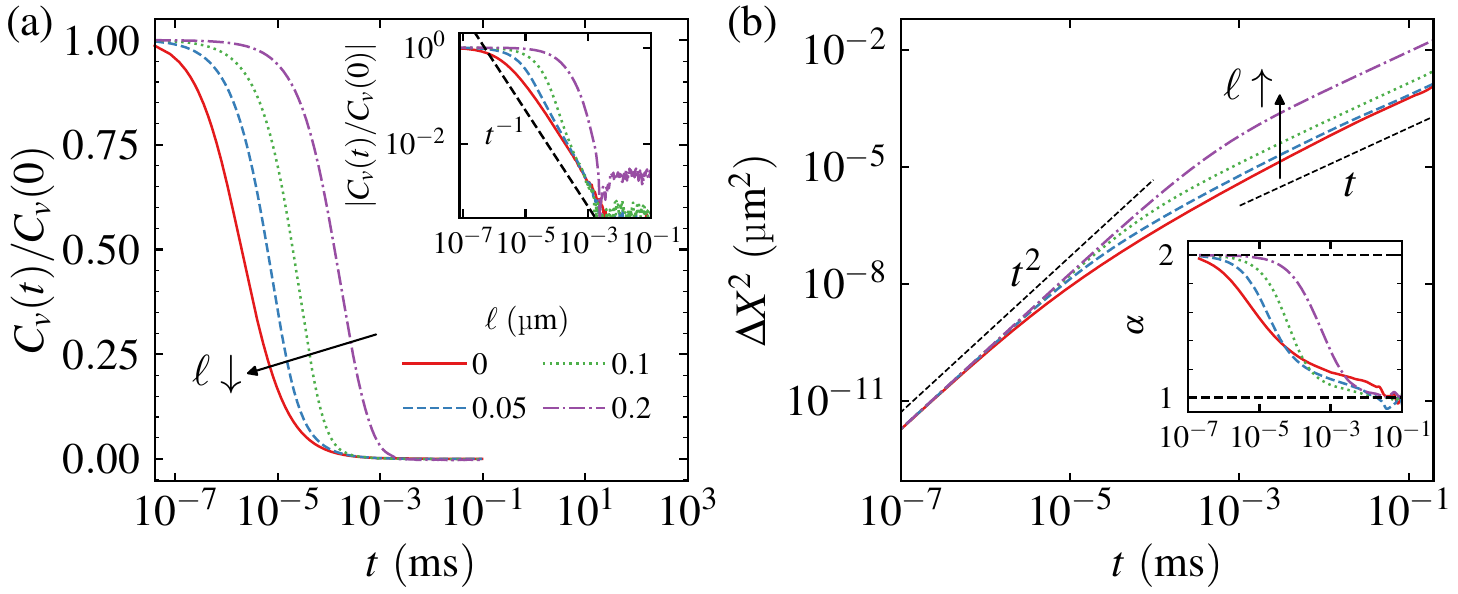}
    \caption{Particle VACF and MSD exhibit a monotonic dependence on the correlation length \(\ell\) for \(\corr_1\). (a) Normalized VACF, and (b) MSD \(\Delta X^2\). Insets in (a) show the absolute value of the VACF on a log--log scale, and in (b) the local slope \(\alpha\) of \(\Delta X^2\). As \(\ell\) increases, the MSD rises, whereas the VACF exhibits an extended ballistic regime. Simulation parameters are the same as \cref{fig:2}.}
    \label{fig:3}
\end{figure}

Representative particle trajectories are shown in \cref{fig:2}(b) for visualization; quantitative analyses follow in \cref{fig:3}. The trajectories reveal that particle diffusion increases monotonically with the correlation length \(\ell\). For small \(\ell\), the trajectory appears irregular and decorrelates rapidly, consistent with Brownian diffusion, whereas larger \(\ell\) produces smoother, more persistent motion characteristic of an extended ballistic regime confirmed in \cref{fig:3}.

To quantify the diffusion dynamics, we compute the particle velocity autocorrelation function (VACF), \(\autocorr(t) = \aab*{\bv(\tau)\bv(\tau+t)}\), and mean-squared displacement (MSD), \(\Delta X^2(t) = \aab*{|\bX(\tau+t) - \bX(\tau)|^2}\)~(\cref{fig:3}), where \(\bv(t)\) and \(\bX(t)\) are the particle velocity and position. For uncorrelated noise (\(\ell=0\)), the VACF decays rapidly at short times and enters the classical \(t^{-1}\) hydrodynamic regime in two dimensions for \(t\gtrsim10^{-5}~\unit{\ms}\)~[\cref{fig:3}(a), inset], reflecting momentum conservation~\cite{alder1970pra,ernst1970prl}.

\begin{figure*}
    \centering    
    \includegraphics[width=\textwidth]{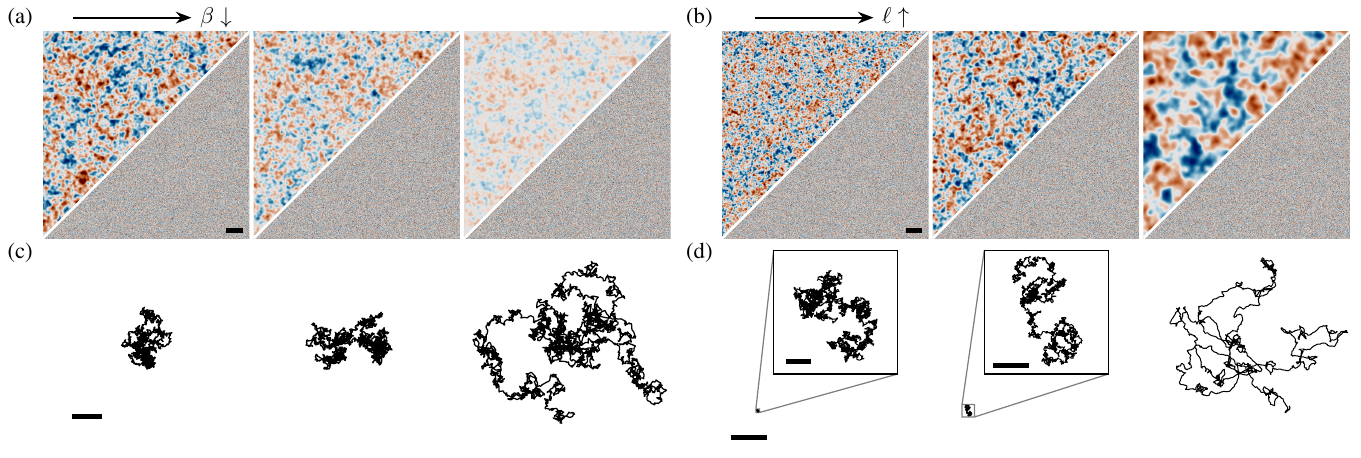}
    \caption{Enhancement of particle diffusion with increasing noise correlation length \(\ell\) increases or decreasing correlation strength \(\beta\) for the correlation function \(\corr_2\). Particle radius \(a=\qty{0.05}{\um}\). Representative instantaneous snapshots of the correlated noise (upper halves) and the resulting fluid velocity (lower halves) for (a) \(\beta=1,0.5\) and \(0.1\) (left to right) at \(\ell=\qty{0.2}{\um}\), and (b) \(\ell=0.1,0.2\) and \qty{0.5}{\um} at \(\beta=1\). Only the \(x\)-components (\(\cZ_{xx}\) and \(u_x\)) are shown, owing to spatial isotropy. Scale bars: \qty{1}{\um}. (c,d) Representative tracer trajectories over the same time span. Scale bars: \(2\times10^{-3}~\unit{\um}\) in (c), \qty{0.02}{\um} in (d), \(4\times10^{-4}~\unit{\um}\) in the inset of (c), and \(2\times10^{-3}~\unit{\um}\) in the inset of (d). Simulation methods and parameters are described in Sec.~II of the Supplemental Material~\citep{supplemental}.}
    \label{fig:4}
\end{figure*}

As \(\ell\) increases, the VACF persists longer, reflecting a prolonged ballistic regime and more persistent particle motion, consistent with the trajectories shown in \cref{fig:2}(b). The dynamics thus become effectively inertia-dominated, even though the particles themselves are inertialess.

The extended ballistic regime is also evident in the MSD~[\cref{fig:3}(b)], which departs from \(t^2\) scaling at progressively later times with increasing \(\ell\). The local slope, \(\alpha(t) = \d\ln\Delta X^2/\d\ln{t}\) [\cref{fig:3}(b), inset], confirms that the ballistic regime (\(\alpha\approx2\)) extends to \(t\lesssim10^{-4}~\unit{\ms}\) for the largest \(\ell\). At longer times, all cases cross over to normal diffusion (\(\alpha=1\)) for \(t\gtrsim10^{-2}~\unit{\ms}\).

\begin{figure}
    \centering
    \includegraphics[width=0.48\textwidth]{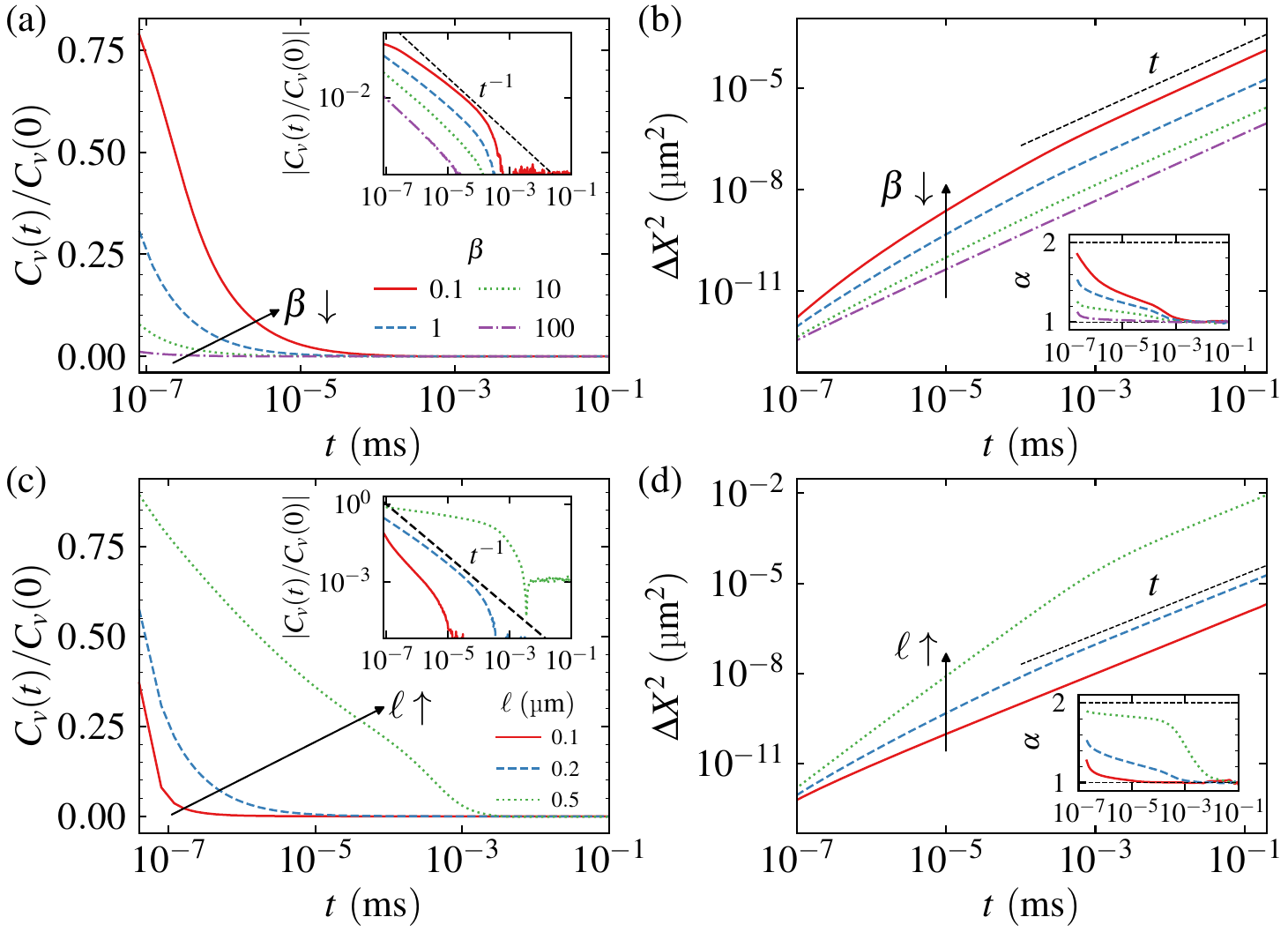}
    \caption{Particle VACF and MSD exhibit a monotonic dependence on the correlation length \(\ell\) and correlation strength \(\beta\) for \(\corr_2\). (a) Normalized VACF, and (b) MSD \(\Delta X^2\) at fixed \(\ell=\qty{0.2}{\um}\). Insets in (a) show the absolute value of the VACF on a log--log scale, and in (b) the local slope \(\alpha\) of \(\Delta X^2\). As \(\beta\) decreases, the MSD rises, whereas the VACF enters a diffusive regime as \(\beta\) increases. (c) Normalized VACF, and (d) MSD \(\Delta X^2\) at fixed \(\beta=1\). As \(\ell\) increases, the MSD rises, and the VACF transitions away from the diffusive regime. Simulation parameters are the same as \cref{fig:4}.}
    \label{fig:5}
\end{figure}

\textit{Suppression of Ballistic Motion}---\Cref{fig:4} compares simulations for the second correlation function, \(\corr_2(r)\), which allows the effects of correlation strength \(\beta\) and length \(\ell\) to be varied independently. \Cref{fig:4}(a,b) show instantaneous fields of the correlated noise and resulting velocity for (a) fixed \(\ell=\qty{0.2}{\um}\) with varying \(\beta\), and (b) fixed \(\beta=1\) with varying \(\ell\). For decreasing \(\beta\), the overall noise amplitude diminishes while spatial structures, set by \(\ell\), remain similar. Conversely, increasing \(\ell\) at fixed \(\beta\) enlarges these structures without changing amplitude. In both cases, the velocity field remains spatially random and consistent with energy equipartition, confirming that the system stays in equilibrium.

Particle trajectories [\cref{fig:4}(b,d)] illustrate the contrasting influence of \(\ell\) and \(\beta\). Particle diffusion strengthens with increasing \(\ell\) or decreasing \(\beta\): larger \(\ell\) or weaker \(\beta\) yields smoother, more persistent motion, while smaller \(\ell\) or stronger \(\beta\) suppresses ballistic excursions and restores nearly diffusive behavior. This complementary dependence of \(\ell\) and \(\beta\) suggests that the enhanced particle diffusion observed for \(\corr_1(r)\) arises from their combined effect.

The corresponding VACF and MSD are shown in \cref{fig:5}. At fixed \(\ell\), increasing \(\beta\) accelerates the decay of the VACF, eliminating the ballistic regime and producing purely diffusive motion for large \(\beta\). At fixed \(\beta\), reducing \(\ell\) has a similar effect, shortening the correlation time and suppressing the ballistic regime. Thus, \(\ell\) and \(\beta\) act as opposing controls over the ballistic regime: larger \(\ell\) enhances, larger \(\beta\) suppresses. Nevertheless, the hydrodynamic long-time tail, \(\autocorr(t)\sim t^{-1}\), remains visible in most cases~[insets in \cref{fig:5}(a,c)].

The MSD~[\cref{fig:5}(b,d)] corroborates these trends: for small \(\beta\) or larger \(\ell\), the initial \(t^2\) growth extends over longer times, whereas for weak large \(\beta\) or small \(\ell\), motion remains nearly diffusive throughout.

\emph{Discussion}---We have derived a spatially correlated fluctuating incompressible Navier--Stokes equation that preserves thermal equilibrium. In this formulation, spatial correlations can either enhance or suppress the ballistic regime, depending on the variations in correlation length and strength. This interplay is encoded in the effective viscosity \(\nueff(r)\), most clearly expressed in Fourier space. Each velocity mode \(\buhat(\bk)\) relaxes with time scale \(\tau_k\sim[\nueffhat(k)k^2]^{-1}\); larger \(\ell\) reduces \(\nueffhat(k)\) at high \(k\), weakening small-scale dissipation and extending the lifetime of short-wavelength modes. Equivalently, in real space, this corresponds to longer-ranged hydrodynamic coupling and slower momentum diffusion. As a result, correlations in the noise field translate into persistent velocity fluctuations and an extended ballistic regime. Conversely, increasing the correlation strength \(\beta\) enhances \(\nueffhat(k)\), and thus viscous dissipation across scales, which shortens relaxation times and suppresses the ballistic regime. 

As the correlation length \(\ell\) increases, the growing relaxation times \(\tau_k\) lead to increasingly sluggish dynamics, reminiscent of the dramatic slowdown observed in glass-forming liquids, where structural relaxation times grow by many orders of magnitude as the temperature approaches the glass transition~\cite{ishino2025natmat,edera2025prx,lois2009prl}. Here, however, the slowdown is governed by spatial scale rather than by time or temperature.

The theoretical framework developed here may be relevant to experimentally accessible systems. Quasi-two-dimensional fluids, such as lipid membranes or thin liquid films, exhibit intrinsic spatial correlations that may play a role analogous to the spatial correlations considered here~\cite{granek2011sm,lavine2002pre,sapp2021pre,schoch2021pnas,fan2010prl}. Such systems may therefore serve as experimental platforms to probe the predicted transition between persistent and overdamped particle dynamics.

\textit{Acknowledgements}---The authors thank members of the Press\'e Lab for helpful discussions. This work was supported by the Army Research Office (ARO) under Grant No.~W911NF-23-1-0304. Numerical simulations were performed on the \textit{Sol} cluster at Arizona State University.

\section{End Matter}

\textit{Fluctuation--Dissipation Relation}---A detailed derivation of the FDR and \cref{eq:generalized_fns} is provided in the Supplemental Material [37]; here we summarize the key steps of the derivation.

The FDR is most conveniently derived in Fourier space by requiring scale-by-scale balance between noise-energy injection and viscous dissipation, as illustrated in~\cref{fig:1}. Since the nonlinear term of the Navier--Stokes equation conserves energy~\cite{bandak2022}, it does not affect the energy balance between noise injection and viscous dissipation that defines the FDR. As a result, the FDR is fully determined by the linearized dynamics (\textit{i.e.}, fluctuating Stokes). Formally, the time evolution of the probability distribution of the Fourier modes \(\buhat(\bk)\) is governed by the Fokker--Planck equation; the FDR ensures that its stationary solution corresponds to thermal equilibrium. Because the Fokker--Planck equations are identical for the nonlinear and linearized forms of the fluctuating Navier--Stokes equation~\cite{bandak2022}, both yield the same equilibrium distribution and hence the same FDR. Therefore, the FDR can be derived directly from the fluctuating Stokes equation, which is a linear system of Langevin equations for \(\buhat\) that can be solved analytically. Solving each Fourier mode yields the FDR
\begin{equation}
    \label{eq:fdr}
    \aab*{\buhat(\bk)\buhat^\dagger(\bk)} = -\nu\Theta\frac{k^2\corrhat(k)}{L(k)}\bP.
\end{equation}
Here, the superscript \(\dagger\) denotes the Hermitian conjugate, \(L(k)\) is the linear diffusion operator (reducing to \(L(k) = -\nu k^2\) for the standard Laplacian), and the projection operator \(\bP = \mathbf{I} - \bk\bk^\top/k^2\), originating from \(\grad p\), enforces the incompressibility condition \(\bk\cdot\buhat(\bk)=0\). Substituting the Laplacian \(L(k) = -\nu k^2\) and \(\corr(r) = \delta(r)\) into \cref{eq:fdr} and taking the trace yields the energy equipartition
\begin{equation}
    \label{eq:mode_energy}
    \aab*{|\buhat(\bk)|^2} = (d-1)\Theta,
\end{equation}
where \(d\) denotes the spatial dimension. The prefactor \(d-1\) follows from \(\mathrm{Tr}(\bP)\), reflecting the removal of the longitudinal mode and its share of thermal energy~\citep{usabiaga2012mms}. 

The diffusion operator consistent with the correlated noise follows from equating the trace of \cref{eq:fdr} to \cref{eq:mode_energy}, yielding 
\begin{equation}
    \label{eq:modified_diffusion_operator}
    L(k) = -\nu\corrhat(k) k^2,
\end{equation}
where we define a wavenumber-dependent effective viscosity as \(\nueffhat(k)=\nu\corrhat(k)\). Replacing the Laplacian diffusion term by \cref{eq:modified_diffusion_operator} and transforming back to real space yields \cref{eq:generalized_fns}.

\begin{figure}
    \centering
    \includegraphics[width=\linewidth]{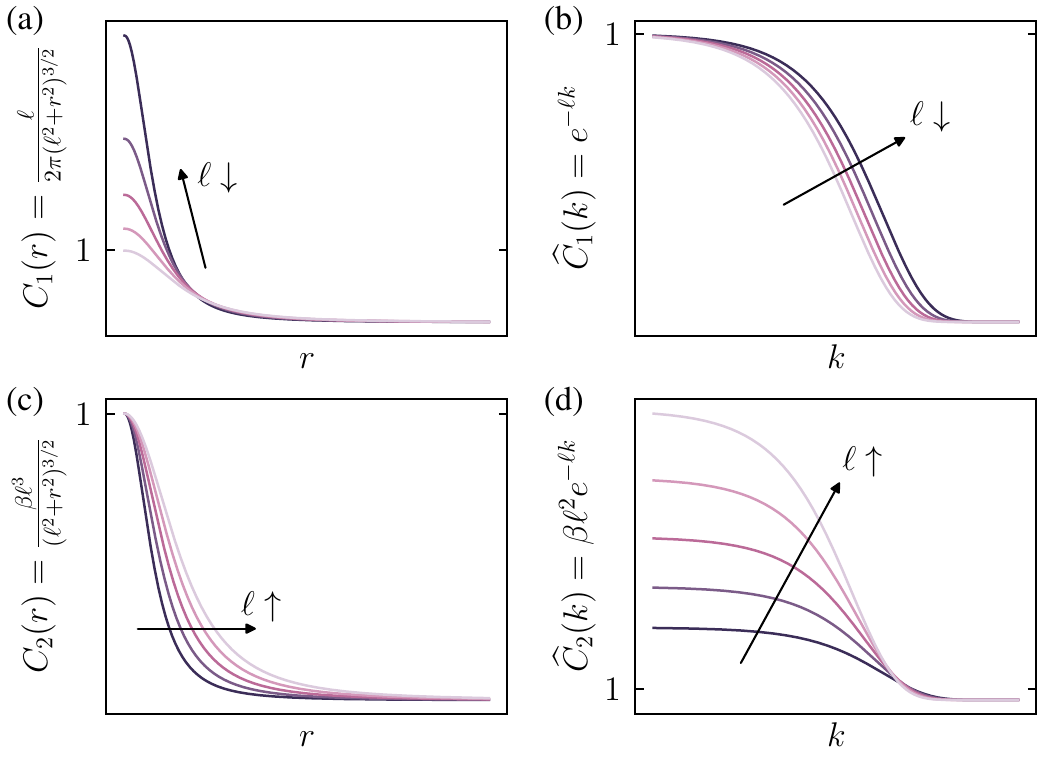}
    \caption{Real- and Fourier-space behavior of the correlation functions \(\corr_1\) and \(\corr_2\). (a) As the correlation length \(\ell\) decreases, \(\corr_1\) narrows toward the delta function. (b) In Fourier space, its spectrum suppresses high-\(k\) contributions. (c) The correlation function \(\corr_2\) maintains a fixed strength \(\corr_2(0)\) while broadening with \(\ell\), allowing independent control of correlation range and strength. (d) Its Fourier spectrum preserves total energy but shifts spectral contributions toward smaller wavenumbers as \(\ell\) increases.}
    \label{fig:6}
\end{figure}

\textit{Correlation Functions}---\Cref{fig:6} illustrates the real- and Fourier-space behaviors of the two correlation functions, \(\corr_1\) and \(\corr_2\), which through the FDR simultaneously modulate the noise amplitude and viscous dissipation, maintaining their scale-by-scale balance.

As \(\ell\) increases, \(\corr_1\) broadens spatially, extending hydrodynamic coupling through the effective viscosity \(\nueff(r)\), and leading to larger-scale structures in the noise field. The reduced correlation strength \(\corr_1(0)\propto\ell^{-2}\) weakens both the noise amplitude and the corresponding viscous dissipation~[\cref{fig:6}(a)].

In Fourier space, \(\corrhat_1(k) = e^{-\ell k}\) yields a \(\nueffhat(k)\) that damps dissipation at large wavenumbers. Consistently, the noise-energy injection spectrum \(k^2\corrhat_1(k)\) decreases with \(k\), reducing small-scale fluctuations and hence the total injected energy~[\cref{fig:6}(b)]. 

The second correlation function \(\corr_2\) fixes the overall amplitude \(\corr_2(0) = \beta\) independently of \(\ell\), enabling separate control of correlation range and strength~[\cref{fig:6}(c)]. In Fourier space, \(\corrhat_2(k)\propto\beta\ell^2e^{-\ell k}\) redistributes both noise-energy injection and viscous dissipation across scales, shifting their contributions toward smaller wavenumbers as \(\ell\) increases~[\cref{fig:6}(d)].

\bibliography{main}

%%%%%%%%%%%%%%%%%%%%%%%%%%%%%%%%%%%%

\end{document}

\begin{CJK*}{UTF8}{gbsn}

\title{
Supplemental Material: Spatially Correlated Noise Induces Transitions from the Diffusive to Ballistic Regime in Fluids
}

\author{Sijie Huang (黄斯杰)}
\author{Ayush Saurabh}
\affiliation{
Department of Physics, Arizona State University, Tempe, AZ 85287, USA
}
\affiliation{
Center for Biological Physics, Arizona State University, Tempe, AZ 85287, USA
}
\author{Steve Press\'e}
\email{Corresponding author: spresse@asu.edu}
\affiliation{
Department of Physics, Arizona State University, Tempe, AZ 85287, USA
}
\affiliation{
Center for Biological Physics, Arizona State University, Tempe, AZ 85287, USA
}
\affiliation{
School of Molecular Sciences, Arizona State University, Tempe, AZ 85287, USA
}

\maketitle

\end{CJK*}

\section{Derivation of the Fluctuation--dissipation relation\label{sec:1}}

In this section, we derive the fluctuation--dissipation relation (FDR) for the spatially correlated fluctuating incompressible Navier--Stokes equation in Eq.~(4) of the main text.

We begin with the original fluctuating incompressible Navier--Stokes equation proposed by \citet{landau1959}
\begin{equation}
    \label{eq:llns}
    \partial_t\bu + \bu\cdot\grad\bu = -\grad p + \nu\nabla^2\bu + \sqrt{2\nu\kBT/\rho}\divergence\bcW, \quad \divergence\bu=0,
\end{equation}
where the random stress tensor \(\bcW\) is a zero-mean, unit-variance Gaussian noise, whose covariance is given by
\begin{equation}
    \label{eq:uncorrelated_noise}
    \aab{\cW_{ij}(\bx,t)\cW_{lm}(\bx',t')} = \pab*{\delta_{il}\delta_{jm} + \delta_{im}\delta_{jl}}\delta(r)\delta(t-t'),
\end{equation}
where \(r=|\bx-\bx'|\) is the spatial separation. In this work, we extend \cref{eq:llns} by replacing \(\bcW\) with a spatially correlated noise \(\bcZ\), whose covariance is defined as 
\begin{equation}
    \label{eq:correlated_noise}
    \aab{\cZ_{ij}(\bx,t)\cZ_{lm}(\bx',t')} = \pab*{\delta_{il}\delta_{jm} + \delta_{im}\delta_{jl}}C(r)\delta(t-t'), 
\end{equation}
where the correlation function \(C(r)\) depends only on \(r\).

After replacing the white noise \(\bcW\) with the spatially correlated noise \(\bcZ\), the corresponding FDR must be derived and enforced to ensure that the system remains in thermal equilibrium. The FDR is most conveniently obtained in Fourier space. In this representation, \cref{eq:llns} with \(\bcW\) replaced by \(\bcZ\) is written as 
\begin{equation}
    \label{eq:fnse_fourier}
    \odv*{\buhat(\bk)}{t} = \bcP\bN(\bk) + L(k)\buhat + \iu\sqrt{2\nu\kBT/\rho}\,\bcP\bk\cdot\bcZhat,
\end{equation}
where $\bN(\bk) = -\widehat{\bu\cdot\grad\bu}$ is the Fourier transform of the nonlinear term, $L(k)$ is the linear diffusion operator, and the projection operator, $\bcP=\mathbf{I}-\bk\bk^\top/k^2$, arising from \(\grad p\), projects \(\buhat(\bk)\) onto the solenoidal subspace to ensure incompressibility \(\bk\cdot\buhat(\bk)=0\). For \cref{eq:fnse_fourier} driven by the uncorrelated noise \(\bcWhat\), \(L(k)\) reduces to the standard Laplacian. In contrast, when the forcing is spatially correlated, the corresponding operator form consistent with the FDR must be determined such that the stationary velocity covariance satisfies equipartition (see below). 

As we already discussed in the main text, the FDR for the fluctuating incompressible Navier--Stokes equation is identical to that of the linearized fluctuating Stokes equation~\citep[see also][]{bandak2022}. In Fourier space, the fluctuating Stokes equation is a linear system of decoupled Langevin equations
\begin{equation}
    \label{eq:fse_fourier}
    \odv*{\buhat}{t}(\bk) = L(k)\buhat + \iu\sqrt{2\nu\kBT/\rho}\,\bcP\bk\cdot\bcZhat.
\end{equation}
This is analogous to the generalized Langevin description of electron--phonon coupling in Ref.~\cite{tamm2018prl}, which likewise incorporates spatially correlated noise. \Cref{eq:fse_fourier} can be solved analytically using integrating factors and It\^o isometry~\cite{oksendal2003}, leading to the covariance of the Fourier modes as 
\begin{equation}
    \label{eq:fourier_mode_autocorr}
    \aab*{\buhat(\bk,t)\buhat^\dagger(\bk,t)} = \aab*{\buhat(\bk,0)\buhat^\dagger(\bk,0)}e^{-L(k)t} - \frac{\nu\kBT}{\rho}\frac{\bcP\bk\aab*{\bcZhat\bcZhat^\dagger}\bk^\dagger\bcP^\dagger}{L(k)}\pab{1 - e^{-2L(k)t}},
\end{equation}
where the superscript $\dagger$ denotes the Hermitian conjugate. By construction, $\bcP$ is real and symmetric, thus $\bcP^\dagger=\bcP$. In the long-time limit, all the exponential terms vanish, yielding
\begin{equation}
    \label{eq:langevin_long_time_limit}
    \aab*{\buhat(\bk,t)\buhat^\dagger(\bk,t)} = -\frac{\nu\kBT}{\rho}\frac{\bcP\bk\aab*{\bcZhat\bcZhat^\dagger}\bk^\dagger\bcP}{L(k)},
\end{equation}
which fixes the equilibrium covariance of the Fourier modes. Substituting the covariance of $\bcZ$ in \cref{eq:correlated_noise} into \cref{eq:langevin_long_time_limit} and carrying out the algebra yields 
\begin{equation}
    \label{eq:covariance}
    \aab*{\buhat(\bk,t)\buhat^\dagger(\bk,t)} = -\frac{\nu\kBT}{\rho}\frac{k^2\corrhat(k)}{L(k)}\bcP,
\end{equation}
\Cref{eq:covariance} provides the FDR consistent with both \cref{eq:fnse_fourier,eq:fse_fourier}, when driven by the correlated noise $\bcZhat$. This expression defines the consistent viscous operator for spatially correlated fluctuations, which is used in all simulations presented below.

For uncorrelated noise ($\ell=0)$, $\corrhat(\bk)$ becomes the delta function, and the diffusion operator reduces to the standard Laplacian $L(k)=-\nu k^{2}$. Inserting these two expressions into \cref{eq:covariance} gives the equilibrium covariance
\begin{equation}
    \label{eq:covariance_simplified}
    \aab*{\buhat(\bk)\buhat^\dagger(\bk)} = \frac{\kBT}{\rho}\bcP.
\end{equation}
The resulting covariance is proportional to the operator $\bcP$, consistent with Ref.~\cite{usabiaga2012mms}. The mean total kinetic energy of the Fourier mode is obtained by taking the trace of the covariance in \cref{eq:covariance_simplified},
\begin{equation}
    \label{eq:equipartition}
    \aab*{|\buhat(\bk)|^2} \equiv \mathrm{Tr}\bab{\aab{\buhat(\bk)\buhat^\dagger(\bk)}} = (d-1)\frac{\kBT}{\rho},
\end{equation}
where $d$ denotes dimensionality. 

Similarly, for correlated noise with $\ell>0$, by taking the trace of \cref{eq:covariance} and invoking the equipartition condition in \cref{eq:equipartition}, we obtain the modified linear diffusion operator corresponding to the correlated noise $\bcZhat$
\begin{equation}
    \label{eq:modified_operator}
    L(k) = -\nu k^2\corrhat(k),
\end{equation}
where \(\nueffhat(\bk) = \nu\corrhat(k)\) is interpreted as the effective viscosity in the main text. \Cref{eq:modified_operator} shows that enforcing equipartition requires modifying the viscous operator, ensuring that dissipation and noise injection remain balanced at each Fourier mode. By the convolution theorem, reverting to real space yields the spatially correlated fluctuating incompressible Navier--Stokes equation
\begin{equation}
    \label{eq:generalized_fnse}
    \partial_t\bu + \bu\cdot\grad\bu = -\grad p + \nu\divergence(\nueff*\grad\bu) + \sqrt{2\nu\kBT/\rho}\,\divergence\bcZ,
\end{equation}
where \(\nueff(r)\) is the inverse Fourier transform of \(\nueffhat(\bk)\).

In discrete formulation, spatially white noise must be treated as a spatiotemporal average since the continuum noise $\bcW(\bx,t)$ is a generalized field and must be represented as a space-time average over each computational cell and time step~\citep[see][and references therein]{usabiaga2012mms}. In the actual numerical schemes, the field $\bcW$ is represented by a set of random numbers $\bm{W}$ according to $(\Delta V_\mathrm{f}\Delta t)^{-1}\bm{W} \leftrightarrow \bcW$, where $\Delta V_{\mathrm{f}}$ is the volume of a computational cell, and $\Delta t$ is the computational timestep. With this averaging, \cref{eq:equipartition} yields the equilibrium energy spectrum for the discrete system
\begin{equation}
    \label{eq:covaraince_discrete}
    \aab*{\buhat(\bk)\buhat^*(\bk)} = (d-1)\frac{\kBT}{\rho\Delta V_\mathrm{f}},
\end{equation}
in agreement with Ref.~\cite{usabiaga2012mms}. In this work, to make the correlated and uncorrelated formulations directly comparable in amplitude, we retain the prefactor \(\Delta V_\mathrm{f}\) for the spatially correlated noise, such that $(\Delta V_\mathrm{f}\Delta t)^{-1}\bm{Z} \leftrightarrow \bcZ$, and the resulting velocity field follows \cref{eq:covaraince_discrete}.

\section{Simulation details}

\subsection{Description of the particle}

Throughout this work, we simulate particle diffusion in 2D fluctuating fluids. This is inherently a fluid--structure interaction problem. To couple the fluid and particle phases, we follow the general strategy of an immersed boundary method developed in Refs.~\cite{usabiaga2013jcp,usabiaga2014cmame}. The method imposes a no-slip condition on the particle surface, enforcing zero relative velocity between the fluid and the particle, thereby conserving the total momentum of the combined system. When the particles are treated as passive and inertialess, they are advected passively by the flow without exerting forces back on the fluid. Their motion reduces to 
\begin{subequations}
    \label{eq:sim_particle}
    \begin{align}
        \odv*{\bX(t)}{t} &= \bv(\bX(t)), \label{eq:sim_particle_motion} \\
        \bv(\bX(t)) &= \bJ(\bX(t))\bu(\bx,t) = \int\delta_a(\bX(t) - \bx)\bu(\bx,t)\d{\bx}, \label{eq:sim_interpolation}
    \end{align}
\end{subequations}
where $\bu(\bx,t)$ is the flow velocity field governed by \cref{eq:generalized_fnse}, $\bX(t)$ is the particle position, $\bv(\bX(t))$ its velocity, and $\delta_a$ an averaging kernel of width $a$ that also defines the particle's volume. The operator $\bJ(\bX(t))$ averages $\bu(\bx)$ around \(\bX(t)\), so that $\bv(\bX(t))$ represents the locally averaged fluid velocity experienced by the particle. Therefore, \cref{eq:sim_particle} describes a passive tracer advected by the coarse-grained flow field \(\bv\)~\cite{donev2014jsm}.

\subsection{Spatiotemporal discretization}

\begin{table}
    \centering
    \renewcommand{\arraystretch}{1.5}
    \begin{tabular}{cc|ccccccccc}
        \hline
        & Case & \(\ell~(\unit{\um})\) & \(\beta\) & $\rho~(\unit{\ug/\um^2})$ & $\nu~(\unit{\um^2/\ms})$ & $L~(\unit{\um})$ & $a~(\unit{\um})$ & $\Delta x~(\unit{\um})$ & $\Delta t~(\unit{\ms})$ & $t_{\text{end}}~(\unit{\ms})$ \\
        \hline
        \multirow{4}{*}{\(\corr_1(r)\)} & 1 & 0 & --- & $10^{-6}$ & $875$ & $4\pi$ & $0.05$ & $0.012$ & $4\times10^{-8}$ & $1$ \\
        & 2 & 0.05 & --- & $10^{-6}$ & $875$ & $4\pi$ & $0.05$ & $0.012$ & $4\times10^{-8}$ & $1$ \\
        & 3 & 0.1 & --- & $10^{-6}$ & $875$ & $4\pi$ & $0.05$ & $0.012$ & $4\times10^{-8}$ & $1$ \\
        & 4 & 0.15 & --- & $10^{-6}$ & $875$ & $4\pi$ & $0.05$ & $0.012$ & $4\times10^{-8}$ & $1$ \\
        \hline 
        \multirow{5}{*}{\(\corr_2(r)\)} & 1 & $0.2$ & 0.1  & $10^{-6}$ & $875$ & $4\pi$ & $0.05$ & $0.012$ & $4\times10^{-8}$ & $0.2$ \\
        & 2 & $0.2$ & 1 & $10^{-6}$ & $875$ & $4\pi$ & $0.05$ & $0.012$ & $4\times10^{-8}$ & $0.2$ \\
        & 3 & $0.2$ & 10 & $10^{-6}$ & $875$ & $4\pi$ & $0.05$ & $0.012$ & $4\times10^{-8}$ & $0.2$ \\
        & 4 & $0.2$ & 100 & $10^{-6}$ & $875$ & $4\pi$ & $0.05$ & $0.012$ & $4\times10^{-8}$ & $0.2$ \\
        \hline 
        \multirow{3}{*}{\(\corr_2(r)\)} & 1 & $0.1$ & 1  & $10^{-6}$ & $875$ & $4\pi$ & $0.05$ & $0.012$ & $4\times10^{-8}$ & $0.2$ \\
        & 2 & $0.2$ & 1 & $10^{-6}$ & $875$ & $4\pi$ & $0.05$ & $0.012$ & $4\times10^{-8}$ & $0.2$ \\
        & 3 & $0.5$ & 1 & $10^{-6}$ & $875$ & $4\pi$ & $0.05$ & $0.012$ & $4\times10^{-8}$ & $0.2$ \\
        \hline
    \end{tabular}
    \caption{Simulation conditions. Fluid is assumed to be water at $T=\qty{300}{\K}$. $\rho$ is the density (in 2D, with units of mass$/$length$^2$), $\nu$ is the kinematic viscosity, $\kBT$ is the thermal energy, $L$ is the domain side length, $a$ is the particle radius, $\Delta x$ is the grid spacing, $\Delta t$ is the simulation timestep and $t_{\text{end}}$ is the simulation time span. \(\ell\) and \(\beta\) are the correlation length and strength, respectively.}
    \label{tab:sim_cond}
\end{table}

We now describe how \cref{eq:generalized_fnse,eq:sim_particle} are solved numerically. \Cref{eq:generalized_fnse} is spatially discretized using a Fourier--Galerkin pseudospectral method with $3/2$-rule de-aliasing to eliminate the aliasing errors arising from the nonlinear term~\cite{canuto2007}. This approach provides spectral accuracy and efficiency, in contrast to the finite-volume discretization used in Refs.~\cite{usabiaga2013jcp,usabiaga2014cmame}. The nonlinear term is advanced in time with Heun's method~\cite{usabiaga2012mms}, while the linear term and the stochastic forcing are treated exactly via integrating factors~\cite{canuto2006,oksendal2003}. Particle trajectories are updated using a midpoint predictor--corrector scheme~\cite{usabiaga2013jcp,usabiaga2014cmame}. The resulting discretized equations are given by 
\begin{subequations}
    \label{eq:discretized_system}
    \begin{align}
        \bX^{\star,n+1} &= \bX^n +\frac{\Delta t}{2}\bJ^n\bu^n, \\
        \buhat^{\star,n+1} &= \Phi\pab{\buhat^n + \Delta t\bcP\bN^{n}} + \iu\sqrt{\Theta}\Gamma\bcP\bk\cdot\widehat{\bm{Z}}^n, \\
        \buhat^{n+1} &= \Phi\buhat^n + \frac{\Delta t}{2}\bcP\pab{\Phi\bN^{n} + \bN^{\star,n+1}} + \iu\sqrt{\Theta}\Gamma\bcP\bk\cdot\widehat{\bm{Z}}^n, \\
        \bX^{n+1} &= \bX^{n} + \Delta t\bJ^{\star,n+1}\pab{\frac{\bu^{n+1} + \bu^n}{2}},
    \end{align}
\end{subequations}
where \(\Delta t\) is the timestep size, 
\[
\Phi(\bk) = e^{\Delta tL(k)},
\quad 
\Gamma(\bk)=\sqrt{\frac{e^{2\Delta tL(k)} - 1}{2L(k)}},
\]
$\bN^n = -\widehat{\bu^n\grad\bu^n}$, $\bJ^n = \bJ(\bX^n)$, and $\Theta=2\nu\kBT/(\rho\Delta V_\mathrm{f}\Delta t)$. The random stress tensor \(\widehat{\bm{Z}}(\bk)\) is sampled according to the discretized normalization introduced in \cref{sec:1}, ensuring that the discrete solver satisfies the same balance between noise energy injection and viscous dissipation as the continuum formulation. To generate this correlated noise efficiently, spatially white noise is multiplied by the prescribed spectral correlation function,
\begin{equation}
    \widehat{\bm{Z}}(\bk) = \corrhat(\bk)\frac{\widehat{\bm{W}} + \widehat{\bm{W}}^\top}{\sqrt{2}},
\end{equation}
which avoids costly real-space convolutions and, through symmetrization, preserves angular momentum~\cite{usabiaga2012mms}. The correlation functions used in this work are
\begin{equation}
    \corr_1(r) = \frac{\ell}{2\pi(\ell^2 + r^2)^{3/2}},\quad \corr_2(r) = \frac{\ell^3}{(\ell^2 + r^2)^{3/2}},
\end{equation}
as described in the main text.

The particle tracking strategy follows Ref.~\cite{donev2014jsm}. The operation $\bJ(\bX(t))$ in \cref{eq:sim_particle} consists of two steps: the fluid velocity field is first low-pass filtered in Fourier space, and then the particle velocity is obtained by interpolating the filtered field at the particle position $\bX(t)$. In the present work, the following Gaussian function is used as the filter kernel to approximate Peskin's four-point kernel~\citep{usabiaga2013jcp}
\begin{equation}
    G_a(\bx) = (\pi a^2)^{-d/2}\exp\pab{-\frac{\bx^2}{a^2}},
\end{equation}
where $a$ is assumed to be the particle radius. The volume of the particle $\Delta V_\mathrm{p}$ can be approximated by $\Delta V_{\mathrm{p}} = \bab{\int G_a^2(\bx)\d{\bx}}^{-1} = (2\pi a^2)^{d/2}$~\cite{usabiaga2013jcp,usabiaga2014cmame}. After filtering, the interpolation is carried out using a GPU-based non-uniform fast Fourier transform~\cite{cufinufft}.

The discretized fluid--particle system described by \cref{eq:discretized_system} is solved in a doubly periodic domain of side length $L=4\pi$~\unit{\um}, discretized on a $1024^2$ uniform grid. The fluid is assumed to be water at the temperature $\qty{300}{\K}$, and all the simulation cases and corresponding parameters are summarized in \cref{tab:sim_cond}. The parameter sets in \cref{tab:sim_cond} span both correlation functions (\(\corr_1(r)\) and \(\corr_2(r)\)) and systematically vary the correlation length \(\ell\) and strength \(\beta\). The particle radius \(a=\qty{0.05}{\um}\). These parameters ensure a clear separation between the system size \(L\) and particle radius \(a\), allowing free diffusion while fully resolving the particle on the grid at a reasonable computational cost. Periodic boundaries mimic an unbounded, homogeneous fluid, eliminating boundary fluxes and ensuring that no energy is injected or removed. They therefore provide the simplest boundary conditions consistent with the FDR, and require no additional boundary corrections~\cite{usabiaga2012mms,delong2014jcp}. The finite domain size primarily influences the long-time behavior of the velocity autocorrelation function (VACF), once momentum has diffused across the system, resulting in an exponential decay in the VACF~\cite{usabiaga2013jchemphys,atzberger2006}. The influence of this finite-size effect is not significant over the short and intermediate timescales that are the focus of this work. All simulations were performed on GPUs using \texttt{CuPy}\footnote{v.13.6.0, \url{https://cupy.dev/}.}.

All simulations are initialized by sampling the fluid velocity field from the equilibrium Boltzmann distribution
\begin{equation}
    \label{eq:boltzmann_fluid}
    P(u_i) = \sqrt{\frac{\Delta m_{\mathrm{f}}}{2\pi\kBT}}\exp\pab{-\frac{\Delta m_{\mathrm{f}}u_i^2}{2\kBT}}, \quad i=1,\,2,
\end{equation}
and then projected onto the solenoidal subspace in Fourier space to enforce incompressibility, \(\divergence\bu=0\). The equilibrium thermal velocity for the fluid is given by $
u_{\mathrm{eq}} = \sqrt{\kBT/\Delta m_{\mathrm{f}}}\approx\qty{166}{\um/\ms}$, where $\Delta m_{\mathrm{f}}=\rho\Delta V_{\mathrm{f}} = \rho\Delta x^2$ denotes the mass of the fluid in one computational cell of grid spacing $\Delta x$. 

\begin{figure}
    \centering
    \includegraphics[width=0.3\linewidth]{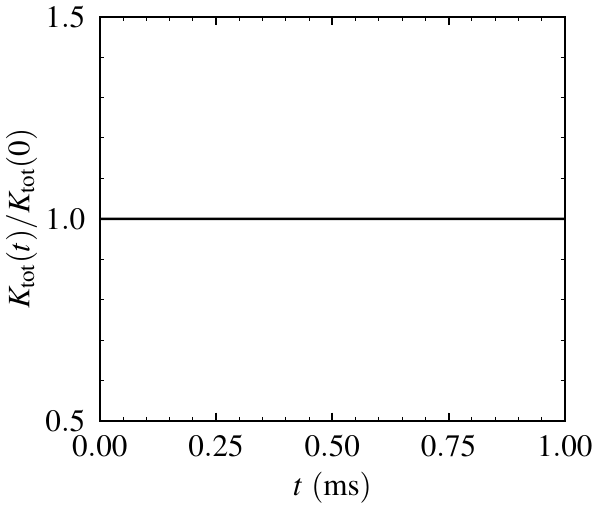}
    \caption{Conservation of total kinetic energy $K_{\text{tot}}(t)$ in the inviscid limit verifies that the nonlinear term is implemented correctly.}
    \label{fig:1}
\end{figure}

Finally, we summarize how the mean-squared displacement (MSD) data are processed to compute the local slope \(\alpha(t)\). In the main text, we present the local slope of the MSD, defined as 
\begin{equation}
    \alpha(t) = \odv{\log\mathrm{MSD}}{\log t}.
\end{equation}
Because the MSD signal is often noisy, we estimate \(\alpha(t)\) using a combination of local averaging and ordinary least-squares (OLS) regression over a sliding window\footnote{We use the \texttt{RollingOLS} function implemented in the \texttt{statsmodel} Python library (v.0.14.5, \url{https://www.statsmodels.org/stable/})}. Specifically, the MSD data are averaged over logarithmically spaced intervals of \(t\), using about 400 bins between the minimum and maximum lag times. In each bin, we compute the geometric mean of the times and the mean MSD value. The resulting pairs, \((\log t_i, \log\mathrm{MSD}(t_i))\), are then fitted by OLS in sliding windows of width 9 bins to obtain a smooth and robust estimate of the local slope \(\alpha(t)\).

\section{Validations}

\subsection{Energy conservation in the inviscid limit}

We first validate our implementation of the nonlinear term. To this end, we consider the inviscid limit, with zero viscosity and noise, where the fluctuating incompressible Navier--Stokes equation reduces to the incompressible Euler equation. In this case, the nonlinear term should conserve the total kinetic energy, 
\[
K_{\text{tot}}(t) = \int_{\Omega}|\bu(\bx,t)|^2\d{\bx},
\]
in the inviscid limit. Here, $\Omega$ denotes the computational domain. Since discretizing this term can introduce spurious energy gain or loss if not handled carefully, we use this property as a stringent test of our numerical implementation. Simulation conditions remain the same as those cases for \(\corr_1(r)\) in \cref{tab:sim_cond}, except that $\nu=0$ and no noise. The resulting energy evolution is shown in \cref{fig:1}, where $K_{\text{tot}}(t)$ remains constant over time. This confirms that the nonlinear term is discretized in an energy-conserving form, and that no spurious energy is injected into the system through discretization errors that would otherwise violate the FDR.

\subsection{Equilibrium statistics and particle diffusion}

\begin{figure}
    \centering
    \includegraphics[width=\linewidth]{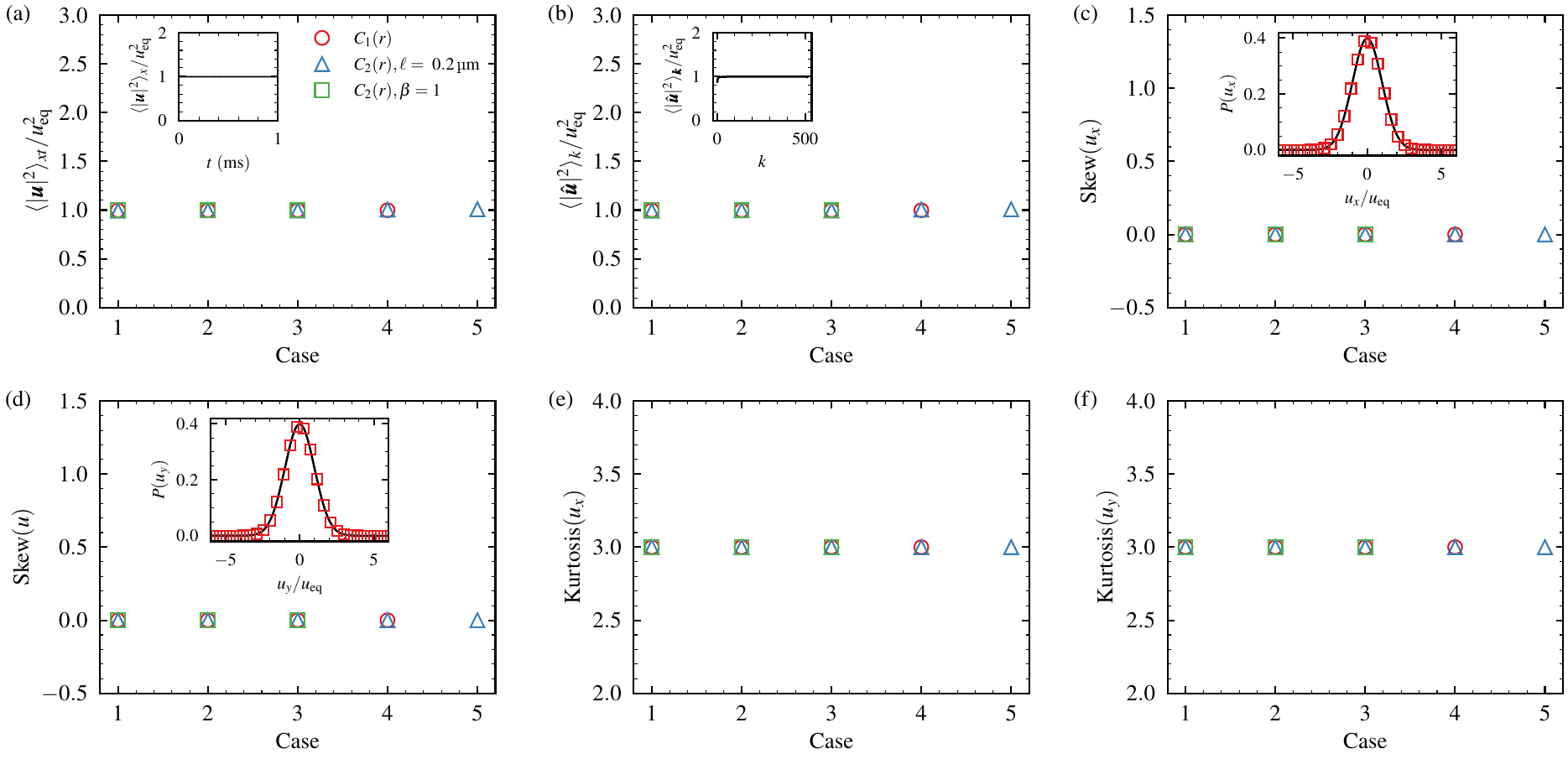}
    \caption{Confirmation that the fluid phase is at thermal equilibrium across all simulations. (a) The spacetime-averaged kinetic energy agrees with its theoretical equilibrium value for all the simulation cases. The inset shows the time evolution of the spatially averaged kinetic energy for a representative case (Case 1 of \(\corr_1(r)\)). (b) Mode-averaged Fourier energy spectra collapse onto the equilibrium value. The inset shows a shell-averaged energy spectrum for a representative case (Case 1 of \(\corr_1(r)\)). (c--f) Skewness and kurtosis of the normalized velocity components, \(u_x/u_{\text{eq}}\) and \(u_yu_{\text{eq}}\), match the values of the Gaussian values (0 and 3), indicating that the velocity field follows the equilibrium Boltzmann distribution. Insets in (c,d) show representative PDFs for Case 1 of \(\corr_1(r)\). Case numbering and simulation parameters are listed in \cref{tab:sim_cond}.}
    \label{fig:2}
\end{figure}

Next, we validate that the numerical solver reproduces the correct equilibrium statistics. We first examine the fluid phase: at thermal equilibrium, the mean kinetic energy fluctuates around the predicted equilibrium value $u_{\mathrm{eq}}^{2}$. \Cref{fig:2}(a) reproduces this behavior with excellent agreement. Satisfaction of the FDR implies energy equipartition among Fourier modes~\cite{delong2013pre,bandak2022,usabiaga2012mms}, predicting a flat energy spectrum at the equilibrium level described by \cref{eq:equipartition}. As shown in \cref{fig:2}(b), the energy spectrum remains flat across all wavenumbers, demonstrating mode-by-mode equipartition of thermal energy. This confirms that our discrete stochastic forcing and viscous operator jointly satisfy the FDR, a nontrivial requirement in fluctuating--hydrodynamics simulations. Moreover, each velocity component should follow the equilibrium Boltzmann distribution \cref{eq:boltzmann_fluid}. \Cref{fig:2}(c,d) confirm this prediction: the measured probability density functions (PDF) coincide with the analytic form. Together, energy conservation, equipartition, and the correct velocity statistics demonstrate that the simulations faithfully reproduce the thermal equilibrium of the fluctuating fluids.

\begin{figure}
    \centering
    \includegraphics[width=\linewidth]{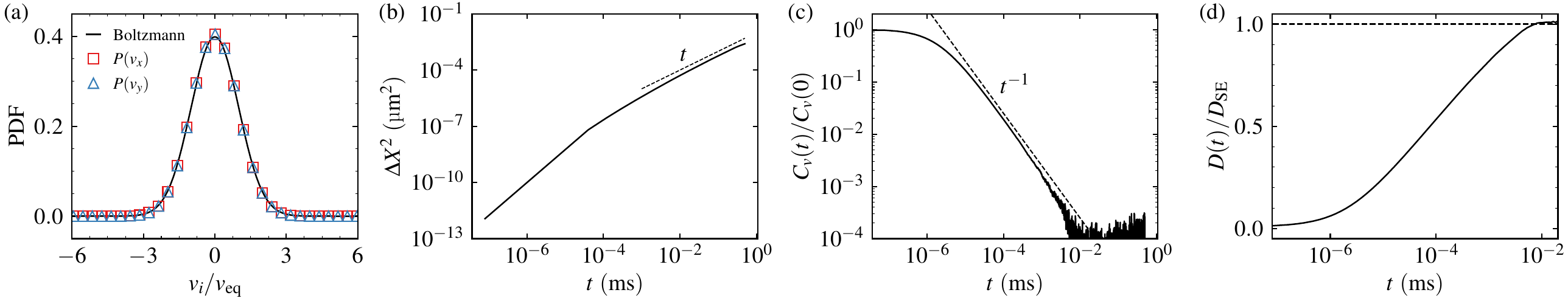}
    \caption{Confirmation of Brownian diffusion in 2D for uncorrelated noise ($\ell=0$). (a) Probability densities of the velocity components $v_x$ and $v_y$ collapse onto the equilibrium Boltzmann distribution. The velocity components are normalized by the corresponding thermal velocity, $v_{\mathrm{eq}}$. (b) MSD ($\Delta X^2$) crossover from ballistic motion ($\Delta X^2\propto t^2$) to normal diffusion ($\Delta X^2\propto t$). (c) VACF exhibits \(t^{-1}\) hydrodynamic tail between $10^{-5}$ and $10^{-2}~\unit{\ms}$, consistent with momentum conservation. (d) Long-time diffusivity (\(t\gtrsim10^{-2}~\unit{\ms}\)) approaches the Stokes--Einstein prediction \(D_{\text{SE}}\) [\cref{eq:stokes_einstein}].}
    \label{fig:3}
\end{figure}

We next validate particle diffusion with uncorrelated noise ($\ell=\qty{0}{\um}$), which should reproduce classic Brownian behavior. The corresponding results are shown in \cref{fig:3}. \Cref{fig:3}(a) presents the measured PDF of the particle velocity, approximated by $\bv^n = (\bX^{n+1} - \bX^n)/\Delta t$~\cite{usabiaga2013jchemphys}, together with the equilibrium Boltzmann distribution
\begin{equation}
    P(v_i) = \sqrt{\frac{\Delta m_\mathrm{p}}{2\pi\kBT}}\exp\pab{-\frac{\Delta m_\mathrm{p}v_i^2}{2\kBT}},\quad i=1,\,2,
\end{equation}
where $\Delta m_{\mathrm{p}}=\rho\Delta V_{\mathrm{p}}$ is the particle mass, and we can define the thermal velocity $v_{\text{eq}} = \sqrt{\kBT/\Delta m_{\mathrm{p}}}$. The agreement confirms that the particles are in thermal equilibrium.

\Cref{fig:3}(b) and (c) repeat the MSD and velocity autocorrelation function (VACF) for the uncorrelated noise from Fig.~4 of the main text to make this validation self-contained. The MSD, $\Delta X^2(t) = \aab*{|\bX(\tau+t) - \bX(\tau)|^2}$, exhibits the expected crossover from ballistic (MSD $\propto t^2$) to normal diffusive motion (MSD $\propto t$). The VACF, $C_v(t) = \aab{\bv(\tau)\bv(\tau+t)}$, displays the hydrodynamic long-time tail $t^{-1}$ between $10^{-5}$ and $10^{-3}~\unit{\ms}$, a consequence of hydrodynamic memory effect due to momentum conservation~\citep{alder1970pra,ernst1970prl}. Beyond this window, finite-domain effects accelerate the decay, consistent with Ref.~\cite{usabiaga2013jchemphys}.

The Stokes--Einstein (SE) relation gives the diffusion coefficient of a particle in 2D as~\citep{donev2014jsm,usabiaga2013jchemphys}
\begin{equation}
    \label{eq:stokes_einstein}
    D_{\mathrm{SE}} = \frac{\kBT}{4\pi\rho\nu}\ln\frac{L}{ca}.
\end{equation}
Here, $c=0.2$ is an empirical constant that depends on the numerical discretization and boundary conditions; its value is obtained from numerical evaluation in our setup (as done also in Ref.~\cite{donev2014jsm}). In the long-time limit, the simulated diffusivity should approach $D_{\mathrm{SE}}$. The time-dependent diffusivity is approximated by~\citep{usabiaga2013jchemphys}
\begin{equation}
    D(n\Delta t) = \frac{\Delta t}{2}C_v(0) + \Delta t\sum_{j=1}^{n-1}\pab{1 - \frac{j}{n}}C_v(j\Delta t).
\end{equation}
\Cref{fig:3}(d) shows that $D(t)$ saturates for $t\gtrsim10^{-2}~\unit{\ms}$, and approaches $D_{\mathrm{SE}}$, confirming the expected diffusion rate of the simulated particle. 

Taken together, the Boltzmann velocity statistics, the ballistic-to-diffusive MSD crossover, the $t^{-1}$ hydrodynamic tail of the VACF, and the saturation of the diffusivity at the SE value---these results confirm that the simulated particle faithfully reproduces 2D Brownian motion with the correct hydrodynamic coupling.

In summary, the validations presented in this section confirm that the numerical solver provides a faithful basis for the correlated-noise simulations discussed in the main text.

\bibliography{main}